# Non-equilibrium statistical mechanics of non-equilibrium damage phenomena, a path approach.

## Short title: Non-equilibrium…

S.G. Abaimov

E-mail: sgabaimov@gmail.com

**Abstract:** In this paper we investigate the applicability of non-equilibrium statistical mechanics to non-equilibrium damage phenomena. As an example, a fiber-bundle model with thermal noise and a fiber-bundle model with decay of fibers are considered. Initially we develop an analogy of the Gibbs formalism for non-equilibrium states. Later we switch from the approach of non-equilibrium states to the approach of non-equilibrium paths. Fluctuating behavior in the system is described in terms of effective temperature parameters. An equation of state and an analog of an energy-balance equation are obtained. Also the formalism of the free energy potential is developed. For fluctuations of paths in the system the statistical distribution is found to be Gaussian and the behavior of the susceptibility is investigated.

**PACS.** 62.20.M- Structural failure of materials - 89.75.-k Complex systems - 05. Statistical physics, thermodynamics, and nonlinear dynamical systems

## 1. Introduction

Damage as a complex phenomenon has been investigated by many studies and a survey of recent developments in damage mechanics can be found in [1-3]. Many attempts [4-11] have been made to apply statistical mechanics to damage phenomena. However, damage phenomena usually exhibit more complex behavior than classical gas-liquid or magnetic systems of statistical mechanics. The reason is that damage has two different forms of appearance. Firstly, in the case of non-thermal systems the occurrence of damage has a complex topological appearance. This behavior can be described by the formalism of equilibrium statistical mechanics [10, 11]. However, all resulting equations in this case are valid not for energy characteristics of damage but for its topological properties. This type of behavior is often observed when the dynamical timescale of fracture is much faster than the timescale of thermal fluctuations and conductivity. In this case *a priori* (quenched) input disorder in a model plays the crucial role.

Secondly, damage behavior inherits thermal (annealed) fluctuations from the media in which it occurs. The main representation of thermal damage fluctuations is that of the Griffith theory. However, contrary to expectations, the application of statistical mechanics here is not straightforward. As we will see below, the formalism of statistical mechanics, to be applicable, again requires switching from the classical energy characteristics of a system to a new type of order parameters.

In many systems these two different forms of damage appearance usually occur simultaneously and make it difficult to investigate possible system's behaviors. Therefore, it would be reasonable to separate these two different types of behavior and study them individually. Later, combined, they can provide extensive explanations for the wide range of physical phenomena. The quenched, topological formalism has been investigated in our previous publications [10, 11]. In this manuscript we make the next step and turn our attention to the non-equilibrium behavior governed by thermal, annealed fluctuations.

In section 2 we introduce models which we will use to illustrate our approach. In section 3 we develop an analogue of the classical Gibbs approach for non-equilibrium states. However, this approach has its restrictions which we discuss in section 4. In section 5 we develop a path' approach, when instead of non-equilibrium states of the system we investigate non-equilibrium paths. First, we define the concepts of a micropath, macropath, and equilibrium path. Later, using these definitions, we map fluctuations of damage on the main concepts of statistical mechanics like temperature, entropy, free energy potential, and an energy-balance equation. However, we find that these concepts are no longer associated with the energy characteristics of the states of the system. Instead, they reveal behavior of other parameters which are topological and dynamical. Also we investigate the behavior of fluctuations for this approach and find 'actual' order parameters which diagonalize the matrix of fluctuations.

## 2. Model

Damage is a complex phenomenon. It can be associated with local and non-local load sharing, brittle and ductile behavior. It can emerge both in one-dimensional and in three-dimensional systems, leading in the latter case to three-dimensional stress patterns of crack formation. The basic principles of damage are often completely disguised by the secondary side effects of its appearance.

Therefore, to investigate the main concepts of behavior, it is reasonable to consider initially a simple model. The basic principles of a new formalism can be illustrated utilizing this model and the approaches developed can later be applied to more complex systems. In this paper as an illustration we consider an annealed fiber-bundle model (further on: FBM). We assume that the number of fibers in the model $N$ is constant and infinite in the thermodynamic limit. Intact fibers all carry the same strain $\varepsilon_f$ which is identically equal to the strain $\varepsilon$ of the total model as a 'black box': $\varepsilon_f \equiv \varepsilon$. In this paper we consider the constant strain $\varepsilon = \mathrm{const}$ as an external boundary constraint of the model. The stress of each intact fiber is assumed to have a linear elastic dependence on the strain until a fiber failure occurs: $\sigma_f = E\varepsilon_f$. The Young modulus $E$ here is assumed to be the same for all fibers. This provides a non-linear stress-strain dependence for the total model although each fiber behaves elastically until its failure occurs. We introduce the 'intact' parameter $L$ as the fraction of intact fibers and the 'damage' parameter $D = 1 - L$ as the fraction of broken fibers. The order parameters in gas-liquid systems are densities of phases; in magnetic systems they are magnetizations of phases. Similarly to this approach we define different phases of damage as phases with different fractions of intact or broken fibers. Both the intact parameter $L$ and the damage parameter $D$ can play the role of an order parameter which distinguishes the phases. Everywhere further we assume that at the initial time $t = 0$ all fibers are intact $L(t=0) = 1$.

The first modification of the FBM we consider is an annealed fiber-bundle model with noise [12-15] (further on: NFBM). Each fiber has thermal fluctuations of its energy characteristics. From statistical mechanics we know that a piston, which works as a gas boundary and is supported by a spring, has oscillations due to the equipartition of energy. In particular, the elastic energy of the spring fluctuates as if a white Gaussian noise would be added to the spring's stress. Similarly we assume that the stress of each fiber $\sigma_f$ has an addition $\Delta\sigma_f$ of a white Gaussian noise with zero mean and standard deviation $\sqrt{k_B T}$. The probability density function of this noise is

$$p(\Delta\sigma_f) = \frac{1}{\sqrt{2\pi k_B T}} e^{-\frac{\Delta\sigma^2}{2k_B T}} \qquad (1)$$

and the cumulative distribution function is

$$P(\Delta\sigma_f) = \frac{1}{\sqrt{2\pi k_B T}} \int_{-\infty}^{\Delta\sigma_f} e^{-\frac{x^2}{2k_B T}} dx. \qquad (2)$$

We consider an ensemble of identical systems. Each system in the ensemble during its evolution realizes some particular noise dependence over its fibers. This particular system does not exhibit any variability. However, different systems in the ensemble fail in different ways. This introduces stochastic fluctuations which only on the ensemble average correspond to $P$. As we will see below, the cumulative distribution function $P$ of the noise is a stochastic constraint (further on: SC $P$) and plays a role similar to the temperature in the canonical ensemble.

Each fiber has an *a priori* assigned strength threshold *s* which we choose to be the same for all fibers and not to change during the model evolution. A fiber can fail only when its stress $\sigma_f = E\varepsilon + \Delta\sigma_f$ exceeds its strength *s*. We consider discrete time-steps *dt* of the model evolution. At each time-step the probability for a fiber to fail is $1 - P(s - E\varepsilon)$ and the probability for a fiber to stay intact is $P(s - E\varepsilon)$. Here we assume that there are no correlations of noise among adjacent fibers. Also, in spite of that the time interval *dt* between the consecutive time-steps is assumed to be small, it is supposed to be much larger than the duration of any noise correlations. Therefore we assume that there are no time correlations of the noise either. Another assumption we will use is that although the time interval between consecutive time-steps has zero limit, the thermodynamic limit of infinite number of fibers $N \rightarrow +\infty$ is taken first. Therefore, although only the small fraction of fibers fails at each time-step, the number of failed fibers is much greater than unity. Although we assume the noise to be Gaussian we will never use this assumption further on in the paper. In fact, we will use only the fact that for the given strain $\varepsilon$ the cumulative distribution function *P* of the noise takes some value $P(s - E\varepsilon)$ which is constant for the model evolution.

The second modification of the FBM we consider (following original terminology [16, 17], a model with 'breaking kinetics') is an annealed decay fiber-bundle model (further on: DFBM). There is no thermal noise in this model. Instead, each fiber has an *a priori* assigned probability $p_f dt$ to fail during the time interval *dt* and a probability $(1 - p_f dt)$ to stay intact where $p_f$ is the decay rate which is constant during the model evolution. We see that we can map the behavior of the DFBM on the behavior of the NFBM if we substitute $(1 - p_f dt)$ in all formulae instead of $P(s - E\varepsilon)$. Therefore further on we will refer only to one of these models, assuming however that all results are also valid and for another model.

## 3. An analogy with the classical Gibbs approach of non-equilibrium states

A FBM during its evolution passes through different microconfigurations of broken and intact fibers. So, for a FBM with $N = 3$ fibers, all possible microconfigurations are |||, ||x, |x|, x||, |xx, x|x, xx|, and xxx where the symbol '|' denotes an intact fiber while the symbol 'x' denotes a broken fiber. Further on in this paper index $n$ will be used to enumerate all possible microconfigurations $\{n\}$ of the FBM. We can unite microconfigurations using some parameters. For example, we can unite all microconfigurations corresponding to the given value of intact parameter $L$ into one macroconfiguration $[L] = \bigcup_{\{n\}:L_{\{n\}}=L} \{n\}$.

Gibbs statistical mechanics identifies microstates of the system with the system's microconfigurations and macrostates of the system with the system's macroconfigurations. For the DFBM we can use the same approach. Let us consider the state of the model at some time $t$. For this time $t$ as a microstate $\{n\}$ we will refer to a microconfiguration $\{n\}$ at this time $t$. We know that the probability for a fiber to fail during time interval $dt$ is $e^{-p_f t} p_f dt$ and the probability for a fiber to fail before the time $t$ is $\int_0^t e^{-p_f t} p_f dt = 1 - e^{-p_f t}$. Therefore, the probability of a microstate $\{n\}$ with $L$ intact and $(1 - L)$ broken fibers at the time $t$ is

$$w_{\{n\}}^{equil}(t,L) = \left(e^{-p_f t}\right)^{NL}\left(1 - e^{-p_f t}\right)^{N(1-L)}.$$ (3)

As a macrostate $[L]$ we refer to a macroconfiguration $[L]$ at the time $t$. The number of microstates corresponding to this macrostate is

$$g_{[L]}(t) = \frac{N!}{(NL)!(N(1-L))!}$$ (4)

and the probability of each of these microstates is given by equation (3). Therefore for the probability of this macrostate we obtain

$$W_{[L]}^{equil}(t,L) = g_{[L]}(t)w_{\{n\}}^{equil}(t,L).$$ (5)

We see that we can map this case on the model of the quenched FBM considered in our previous publications [10, 11]. For this we need to substitute $1 - e^{-p_f t}$ instead of $P(E\varepsilon)$ in [10, 11]. For details we refer a reader to these studies. Here we present only the results that immediately follow from them:

$L^{(0)}(t) = e^{-p_f t}$ for the equation of state;

$S_{[L]}(t) = \ln g_{[L]}(t)$ for the entropy of a macrostate;

and $S^{equil}(t) \approx \ln g_{[L^{(0)}]}(t) \approx -Ne^{-p_f t}\ln e^{-p_f t} - N(1 - e^{-p_f t})\ln(1 - e^{-p_f t})$ for the entropy of the ensemble at the time $t$.

The distribution of probabilities (3) we can rewrite as

$$w_{\{n\}}^{equil}(t,L) = \frac{1}{Z(t)}e^{-NL/T(t)}$$ (6)

where $T(t) = \ln^{-1}\left(e^{p_f t} - 1\right)$ is the topological temperature prescribed to the ensemble and $Z(t) = \left(1 - e^{-p_f t}\right)^{-N}$ is the partition function. Then for the Helmholtz energy we have $A_{[L]} = -T\ln\left(ZW_{[L]}^{equil}\right)$ for the macrostate $[L]$ and $A^{equil} = -T\ln Z$ for the ensemble. For an analogue of the energy-balance equation we have the equation of topological balance $N\delta L^{(0)}(t) = T(t)\delta S^{equil}(t)$ where differentials are not with respect to the change of time but for the ensemble variation with respect to the variation of the stochastic constraint $p_f$.

The fluctuations are Gaussian

$$W_{[L]}^{equil}(t, L) \propto \exp\left(-\frac{\Delta L^2}{2L^{(0)}(t)(1 - L^{(0)}(t))/N}\right) \tag{7}$$

and the susceptibility (or the specific heat) is

$$\chi(t) \propto \left\langle\left(NL(t) - N\langle L(t)\rangle^{equil}\right)^2\right\rangle^{equil} = \frac{\partial^2 \ln Z(t)}{\partial(1/T(t))^2} = -\frac{\partial N\langle L(t)\rangle^{equil}}{\partial(1/T(t))} \propto NL^{(0)}(t)(1 - L^{(0)}(t)). \tag{8}$$

## 4. Restrictions of the classical Gibbs approach for states

In the previous section we were lucky that our system permitted the analytical solution. This was the consequence of the fact that we were able *a priori* to find the probability $\int_0^t e^{-p_f t} p_f \, dt = 1 - e^{-p_f t}$ for a fiber to fail before the time $t$. For the case of an arbitrary thermal system it is generally not so. For the general case we know the distribution of probabilities only for a one time-step. In other words, if at time $t_i$ a system in the ensemble is in a microstate $\{n_i\}$ with $NL_i$ intact fibers and probability $w_{\{n_i\}}^{equil}$, we know that the probability for this system at the next time-step $t_{i+1}$ to be at a microstate $\{n_{i+1}\}$ with $NL_{i+1}$ intact fibers is

$$w^{equil}_{\{n_{i+1}\}} = w^{equil}_{\{n_i\}} \left(1 - p_f t\right)^{NL_{i+1}} \left(p_f t\right)^{N(L_i - L_{i+1})}. \tag{9}$$

More general, we utilized here the general Gibbs formula

$$\frac{dw^{equil}_{\{\}}}{dt}(t) = F\left(w^{equil}_{\{\}}(t)\right) \tag{10}$$

that the evolution of the distribution of probabilities of microstates is determined by some functional dependence on the previous history of this evolution. We assumed in equation (10) that the process is a Markov process of order 1 and the evolution depends only on the current state of the system at this time $t$. Being even more general, we would have to include the functional dependence on the total evolution before the time $t$

$$\frac{dw^{equil}_{\{\}}}{dt}(t) = F\left[w^{equil}_{\{\}}(t'), t' \le t\right]. \tag{11}$$

For a system of theoretical mechanics this equation would be

$$\frac{dw^{equil}_{\{\}}}{dt}(t) = \left[H, w^{equil}_{\{\}}(t)\right], \tag{12}$$

where $H$ is a Hamiltonian of this system. For a quantum system we similarly have

$$i\hbar \frac{d\tilde{w}^{equil}_{\{\}}}{dt}(t) = \left[\tilde{H}, \tilde{w}^{equil}_{\{\}}(t)\right]. \tag{13}$$

For other systems we have equation (10) prescribed by the rules of system evolution.

To obtain the distribution of probabilities for microstates at time $t$ we have to integrate equation (10) over all possible paths among configurations. Any microconfiguration at the time $t$ is a result of a tremendous number of different paths leading to this microconfiguration. We have to integrate all these combinatorial paths with their own probabilities to obtain the final distribution of probabilities for microstates at the time $t$. The probabilities of paths depend on all intermediate configurations and not only on the final microstate. Also some of the paths can be prohibited (for the FBM a broken fiber cannot become intact again). Therefore the integration of combinatorial sums becomes cumbersome.

This approach corresponds to the classical Gibbs non-equilibrium statistical mechanics, when microstates of a system are identified with the system's microconfigurations and macrostates of a system are identified with the system's macroconfigurations. However, we know that the probabilities for Markov processes are associated not with the states but with the paths among these states. Therefore it is much easier to find a distribution of probabilities for a macrogroup of paths than for all paths leading to a macrostate. In Gibbs equation (10) the states of a system were chosen as bases, although we see that everything points on the fact that as bases we should choose not the states but the paths. In the next section we will see how to develop an approach, different from Gibbs' one, associated not with the system's states but with the system's paths. The benefit of this approach will be that its bases will not include integrals of the system states for the previous system evolution but, in contrast, will be this evolution itself.

## 5. Statistical mechanics of path approach

In the previous sections and our previous publications [10, 11] we identified microstates of a system with the system's configurations and macrostates of a system with the system's macroconfigurations. However, this approach worked well only for equilibrium statistical mechanics. For non-equilibrium statistical mechanics we should develop another approach.

As an illustration, we consider the case of the NFBM and assume that the process consists of $\nu$ time intervals of duration $dt$, from $t_0 = 0$ to $t_\nu = \nu \cdot dt$. For each time-step $t_i$ as an order parameter we choose the value of the intact parameter $L_i$ at this time-step. For the total process the order parameter is the total history of the intact parameter $L(t) \equiv L_0, \ldots, L_\nu$. We assume that the process always starts from zero damage $L_0 = 1$, so the quantity $L_0$ will not be variable in the ensemble. We assume that broken fibers cannot become intact again, therefore increments of the intact parameter $\Delta L_i \equiv L_i - L_{i-1}$ are always negative.

At each time-step $t_i$ a particular system in the ensemble has own value of intact parameter $L_i$ and is in one of microconfigurations $\{n_i\}$ corresponding to this intact parameter. The next possible microconfiguration for this system at the time-step $t_{i+1}$ can be only a microconfiguration in which all broken fibers remain broken. This makes our system a Markov process of order 1. For the total process from $t_0 = 0$ to $t_\nu = \nu \cdot dt$ we construct all possible chains of microconfigurations. Each such chain as a possible sequence of particular microconfigurations $\{n_0\}, \{n_1\}, \ldots, \{n_\nu\}$ will be referred to as a micropath $\{n_0\} \to \{n_1\} \to \ldots \to \{n_v\}$ (further on we will abbreviate this notation as $\{n_0\} \to \{n_v\}$). For example, one of possible micropaths is a path when the fiber number one fails at the first time-step, the fiber number two fails at the second time-step, and so on.

Let's assume that for a micropath $\{n_0\} \to \{n_v\}$ the sequence of the configurations has the evolution of the intact parameter $L(t) \equiv L_0, \ldots, L_v$. Then the probability of this micropath is

$$w^{equil}_{\{n_0\} \to \{n_v\}}[L(t)] = (1-P)^{N|\Delta L_1|} P^{N(L_0 - |\Delta L_1|)} \ldots (1-P)^{N|\Delta L_v|} P^{N(L_{v-1} - |\Delta L_v|)} = \left(\frac{1-P}{P}\right)^{N(1-L_v)} P^{N \sum_{i=1}^{v} L_{i-1}} \qquad (14)$$

as the probability for $N|\Delta L_i|$ fibers to fail and for $NL_i \equiv N(L_{i-1} - |\Delta L_i|)$ fibers to stay intact, where $P \equiv P(s - E\varepsilon)$ is constant. This probability $w^{equil}_{\{n_0\} \to \{n_v\}}$ is dictated by the prescribed SC $P$. This SC is a model input and acts similarly to the temperature prescribed in the canonical ensemble. An external medium dictates the equilibrium distribution of probabilities for different paths but a system actually can realize a non-equilibrium probability distribution $w_{\{n_0\} \to \{n_v\}}$ for paths. Only the equilibrium distribution of probabilities is dictated by the SC $P$; therefore we used abbreviation '*equil*' to emphasize that this probability distribution corresponds to the equilibrium with the SC $P$.

As to a macropath $[L_0] \to [L_1] \to \ldots \to [L_v]$ (further on we will abbreviate this notation as $[L_0] \to [L_v]$) we will refer to a subset of all micropaths $\{n_0\} \to \{n_v\}$ corresponding to the specified evolution of the intact parameter $L(t)$:

$[L_0] \to [L_\nu] = \bigcup\limits_{\{n_0\} \to \{n_\nu\}:L_{\{n_i\}}=L_i} \{n_0\} \to \{n_\nu\}$. The probabilities of these micropaths are given by

equation (14) and the number of these micropaths is

$$g_{[L_0]\to[L_\nu]} = \frac{(NL_0)!}{(N\mid\Delta L_1\mid)!(N(L_0-\mid\Delta L_1\mid))!} \cdot \frac{(NL_1)!}{(N\mid\Delta L_2\mid)!(N(L_1-\mid\Delta L_2\mid))!} \cdot \ldots \cdot \frac{(NL_{\nu-1})!}{(N\mid\Delta L_\nu\mid)!(N(L_{\nu-1}-\mid\Delta L_\nu\mid))!}$$

$$(15)$$

as a combinatorial choice of $N|\Delta L_i|$ failed fibers among $NL_{i-1}$ initial fibers. We can cancel $(NL_i)!$ in numerators and $(N(L_{i-1} - |\Delta L_i|))!$ in denominators to obtain

$$g_{[L_0]\to[L_\nu]} = \frac{(NL_0)!}{(NL_\nu)!\prod\limits_{i=1}^{\nu}(N\mid\Delta L_i\mid)!} \approx_{\ln} (L_\nu)^{-NL_\nu}\prod\limits_{i=1}^{\nu}\mid\Delta L_i\mid^{-N\mid\Delta L_i\mid}, \qquad (16)$$

where symbol "$\approx_{\ln}$" means that in the thermodynamic limit $N \to +\infty$ all power-law multipliers are neglected in comparison with the exponential dependence on $N$. Everywhere further on, the symbol "$\approx_{\ln}$" will mean the accuracy of an exponential dependence on $N$ neglecting all power-law dependences. For the logarithm of such equations we will use symbol "$\approx$".

The probability for the system to have a macropath $[L_0] \to [L_\nu]$ (to move among macroconfigurations with the specified $L(t)$) is

$$W_{[L_0]\to[L_\nu]}^{equil}[L(t)] = \sum\limits_{n=1}^{g_{[L_0]\to[L_\nu]}} w_{\{n_0\}\to\{n_\nu\}}^{equil}[L(t)] = g_{[L_0]\to[L_\nu]}w_{\{n_0\}\to\{n_\nu\}}^{equil}[L(t)]. \qquad (17)$$

We used above the term 'equilibrium' but did not specify what we refer to using this term. The wrong way would be to imagine a system in some detailed balance. We study the non-equilibrium evolution of the system far from the equilibrium state. Using the term 'equilibrium' we refer to the ensemble of paths whose stochastic properties correspond to the prescribed SC $P$ (whose stochastic properties are in equilibrium with the prescribed SC $P$). In other words, if the ensemble of systems chooses its paths in accordance with equation (14), we will refer to these paths as being in equilibrium with the prescribed SC $P$. However, we also can consider other ensembles which do not obey the prescribed SC $P$ and follow in their evolutions some non-equilibrium distributions of probabilities $w_{\{n_0\}\to\{n_v\}}$ for paths. These ensembles we will refer to as non-equilibrium.

For the equilibrium we will use two different definitions. The SC $P$ is assumed to prescribe the equilibrium probability distribution $w_{\{n_i\}\to\{n_{i+1}\}}^{equil}$ for all micropaths. Therefore, the equilibrium with this SC could be identified with an ensemble of systems which realizes all micropaths with equilibrium probabilities (14): $w_{\{n_0\}\to\{n_v\}} = w_{\{n_0\}\to\{n_v\}}^{equil}$. In other words, all micropaths are possible but their probabilities are dictated by the SC $P$. The superscript '*equil*' will be used for this definition. Then the value of any time-dependent quantity $f(t)$ in equilibrium with the SC $P$ is by definition

$$\left\langle f(t) \right\rangle^{equil} \equiv \sum_{\{n_0\}\to\{n_v\}} w_{\{n_0\}\to\{n_v\}}^{equil} f_{\{n_0\}\to\{n_v\}}(t). \tag{18}$$

In contrast, another definition of the equilibrium is the equilibrium (most probable) macropath, *i.e.,* an ensemble that realizes only (that is isolated in) a subset of micropaths corresponding to the most probable macropath. This is the macropath which gives the main contribution to the partition function. In other words, this is the ensemble which follows only those macropaths $[L_0] \rightarrow [L_\nu]$ which correspond to the maximum of $W^{equil}_{[L_0] \rightarrow [L_\nu]}[L(t)]$ in the space of all possible functions $L(t)$. To distinguish this case the superscript '(0)' will be used.

As an example, we may consider the equilibrium time dependence of the intact parameter $L(t)$. As $\langle L(t) \rangle^{equil}$ we refer to the intact parameter evolution averaged over the equilibrium distribution of probabilities

$$\langle L(t) \rangle^{equil} \equiv \sum_{\{n_0\} \rightarrow \{n_\nu\}} L_{\{n_0\} \rightarrow \{n_\nu\}}(t) w^{equil}_{\{n_0\} \rightarrow \{n_\nu\}} \equiv$$

(19)

$$\equiv \begin{cases} L_0, \, t = t_0 \\ \sum_{\{n_0\} \rightarrow \{n_\nu\}} L_{\{n_i\}} w^{equil}_{\{n_0\} \rightarrow \{n_\nu\}}, \, t = t_i \end{cases} = \begin{cases} L_0, \, t = t_0 \\ \sum_{[L_0] \rightarrow [L_\nu]} g_{[L_0] \rightarrow [L_\nu]} L_i w^{equil}_{\{n_0\} \rightarrow \{n_\nu\}}, \, t = t_i \end{cases}.$$

As $L^{(0)}(t)$ we refer to the dependence of the intact parameter corresponding to the most probable macropath:

$$W^{equil}_{[L_0^{(0)}] \rightarrow [L_{i+1}^{(0)}]}[L^{(0)}(t)] = \max_{L_1, \dots, L_\nu} W^{equil}_{[L_0] \rightarrow [L_\nu]}[L(t)].$$

(20)

Of course, in the thermodynamic limit these quantities are equal: $\langle L(t) \rangle^{equil} \approx L^{(0)}(t)$.

Now we consider a system isolated in a macropath $[L_0] \rightarrow [L_\nu]$. The number $g_{[L_0] \rightarrow [L_\nu]}$ of micropaths corresponding to this macropath is given by equation (16) and the probability of any of these microstates is $w_{\{n_0\} \rightarrow \{n_\nu\}} = 1/g_{[L_0] \rightarrow [L_\nu]}$ (because the system is isolated in this macropath). Because the criterion of isolation in a macropath is not in equilibrium with the SC $P$, the probability obtained does not correspond to the equilibrium distribution for paths (14) and we have not used the superscript 'equil'. The entropy of this macropath is

$$S_{[L_0] \to [L_\nu]} \equiv - \sum_{\{n_0\} \to \{n_\nu\}: L_{\{n_i\}} = L_i} w_{\{n_0\} \to \{n_\nu\}} \ln w_{\{n_0\} \to \{n_\nu\}} = \ln g_{[L_0] \to [L_\nu]}. \tag{21}$$

We should emphasize here that so introduced entropy is the 'dynamical' entropy of the distribution of probabilities for the paths and must not be associated with the classical Gibbs entropy associated with the distributions of probabilities for the states (configurations). In our notations the classical Gibbs entropy would be $S_{\{n\}}(t) \equiv - \sum_{\{n\}} w_{\{n\}}(t) \ln w_{\{n\}}(t)$ as the average at time $t$ over the probabilities $w_{\{n\}}(t)$ of the microconfigurations at this time. Our entropy $S_{\{n_0\} \to \{n_\nu\}} \equiv - \sum_{\{n_0\} \to \{n_\nu\}} w_{\{n_0\} \to \{n_\nu\}} \ln w_{\{n_0\} \to \{n_\nu\}}$ is associated with the probabilities $w_{\{n_0\} \to \{n_\nu\}}$ of micropaths of the total process and cannot be attributed to system characteristics at a particular time $t$.

For the equilibrium with the SC $P$ the distribution of probabilities is the equilibrium distribution (14): $w_{\{n_0\} \to \{n_\nu\}} = w_{\{n_0\} \to \{n_\nu\}}^{equil}$. Therefore the equilibrium entropy is

$$S^{equil} \equiv - \sum_{\{n_0\} \to \{n_\nu\}} w_{\{n_0\} \to \{n_\nu\}}^{equil} \ln w_{\{n_0\} \to \{n_\nu\}}^{equil} = - \sum_{[L_0] \to [L_\nu]} g_{[L_0] \to [L_\nu]} w_{\{n_0\} \to \{n_\nu\}}^{equil} \ln w_{\{n_0\} \to \{n_\nu\}}^{equil}. \tag{22}$$

The function $W_{[L_0] \to [L_\nu]}^{equil}$, given by equation (17), is a product of $g_{[L_0] \to [L_\nu]}$ and $w_{\{n_0\} \to \{n_\nu\}}^{equil}$. Both these functions contain an exponential dependence on $N$ ($N$ is infinite in the thermodynamic limit). Therefore the function $W_{[L_0] \to [L_\nu]}^{equil}$ has a very narrow maximum at the most probable, equilibrium macropath $[L_0^{(0)}] \to [L_\nu^{(0)}]$. As we will see below, the width of this maximum is proportional to $1/\sqrt{N}$. The number of different macropaths $[L_0] \to [L_\nu]$ in the width of this maximum has a power-law dependence on $N$ while the number $g_{[L_0] \to [L_\nu]}$ of micropaths $\{n_0\} \to \{n_\nu\}$ corresponding to each of these macropaths $[L_0] \to [L_\nu]$ has the exponential dependence on $N$. For the normalization of the function $W^{equil}_{[L_0] \to [L_\nu]}$ with the logarithmic accuracy we obtain

$$1 = \sum_{[L_0] \to [L_\nu]} W^{equil}_{[L_0] \to [L_\nu]} \approx_{\ln} W^{equil}_{[L_0^{(0)}] \to [L_\nu^{(0)}]} \equiv g_{[L_0^{(0)}] \to [L_\nu^{(0)}]} w^{equil}_{\{n_0\} \to \{n_\nu\}}[L^{(0)}(t)]. \tag{23}$$

Therefore we can conclude that

$$g_{[L_0^{(0)}] \to [L_\nu^{(0)}]} \approx_{\ln} 1/w^{equil}_{\{n_0\} \to \{n_\nu\}}[L^{(0)}(t)]. \tag{24}$$

In equation (22) for the equilibrium entropy the function $\ln w^{equil}_{\{n_0\} \to \{n_\nu\}}$ has a power-law dependence on $N$ in comparison with the functions $g_{[L_0] \to [L_\nu]}$ and $w^{equil}_{\{n_0\} \to \{n_\nu\}}$ which have the exponential dependences on $N$. Therefore for the equilibrium entropy we obtain

$$S^{equil} \approx -\ln w^{equil}_{\{n_0\} \to \{n_\nu\}}[L^{(0)}(t)] \approx \ln g_{[L_0^{(0)}] \to [L_\nu^{(0)}]} = S_{[L_0^{(0)}] \to [L_\nu^{(0)}]}. \tag{25}$$

We can rewrite the equilibrium distribution of probabilities, given by equation (14), as

$$w^{equil}_{\{n_0\} \to \{n_\nu\}}[L(t)] = \frac{1}{Z^{equil}} \exp\left(-NL_\nu/T_\nu - N\left(\sum_{i=1}^{\nu} L_{i-1}\right)/T\right) \tag{26}$$

where $Z^{equil} = \left(\dfrac{P}{1-P}\right)^N$, $T_\nu = \ln^{-1}\{(1-P)/P\}$, and $T = -\ln^{-1} P$. It is easy to see that $Z^{equil}$ is the path partition function of the system $Z^{equil} = \sum_{\{n_0\} \to \{n_\nu\}} \exp\left\{-NL_\nu/T_\nu - N\left(\sum_{i=1}^{\nu} L_{i-1}\right)/T\right\}$. $T_{L_\nu}$

and $T$ play the roles of the temperatures. We see that when $P$ is close to unity (the probability for a fiber to fail is small) the temperature $T$ is infinite but positive. We should mention that if we would choose the damage parameter $D$ as an order parameter instead of $L$, the temperature would be equal to $T = \ln^{-1} P$ and negative, which, of course, is only due to the change of the sign of the order parameter.

The definition of the temperature $T_\nu$ could, for the first glance, look fictitious. Indeed, we can rewrite equation (14) for probabilities as

$$w_{\{n_0\} \to \{n_\nu\}}^{equil}[L(t)] = (1-P)^{N|\Delta L_1|} P^{NL_1} ... (1-P)^{N|\Delta L_\nu|} P^{NL_\nu} = (1-P)^{N(1-L_\nu)} P^{N \sum_{i=1}^{\nu} L_i} . \qquad (27)$$

This gives $T_\nu = \ln^{-1}(1-P)$ which is different from the expression $T_\nu = \ln^{-1}\{(1-P)/P\}$ above. However, the difference is $\ln(P)$ which is negligible in comparison with $\ln(1-P)$ in the limit $P \to 1$.

The temperature $T = -\ln^{-1} P$ of the system is complementary to the integral of damage evolution: $N \sum_{i=1}^{\nu} L_{i-1} \propto N \int_0^{t_\nu} L(t)dt$. If at the final time-step $t_\nu$ the total system fails: $L(t_\nu) = 0$, then the single order parameter left is the integral $N \int_0^{t_\nu} L(t)dt$. To the extent of our knowledge, this is the first research that points out the importance of the integral of damage evolution for non-equilibrium damage phenomena.

For the system isolated in a macropath $[L_0] \to [L_\nu]$ the probabilities of micropaths are $w_{\{n_0\} \to \{n_\nu\}} = 1/g_{[L_0] \to [L_\nu]}$ and we define the Helmholtz energy of this macropath as

$$A_{[L_0]\to[L_\nu]} \equiv NL_\nu/T_\nu + N\left(\sum_{i=1}^{\nu} L_{i-1}\right)/T - S_{[L_0]\to[L_\nu]} =$$

(28)

$$= -\ln\left\{ g_{[L_0]\to[L_\nu]} \exp\left( -NL_\nu/T_\nu - N\left(\sum_{i=1}^{\nu} L_{i-1}\right)/T \right)\right\} = -\ln Z^{equil}_{[L_0]\to[L_\nu]},$$

where $Z^{equil}_{[L_0]\to[L_\nu]}$ is the partial path partition function [11] of this macropath:

$$Z^{equil}_{[L_0]\to[L_\nu]} = \sum_{\{n_0\}\to\{n_\nu\}\in[L_0]\to[L_\nu]} \exp\left\{ -NL_\nu/T_\nu - N\left(\sum_{i=1}^{\nu} L_{i-1}\right)/T \right\} = Z^{equil}W^{equil}_{[L_0]\to[L_\nu]}.$$ Therefore for the

Helmholtz energy we obtain $A_{[L_0]\to[L_\nu]} = -\ln\left(Z^{equil}W^{equil}_{[L_0]\to[L_\nu]}\right)$.

A careful Reader notices that in contrast to the classical definition $A_{[L_0]\to[L_\nu]} = -T\ln Z^{equil}_{[L_0]\to[L_\nu]}$ we did not multiplied the logarithm of the partial partition function by the temperature. Gas-liquid systems at constant temperature $T = \text{const}$, constant pressure $P = \text{const}$, and constant chemical potential $\mu = \text{const}$ have three 'effective' temperatures: $T$, $T/P$, and $-T/\mu$, and one of them (which is $T$) is chosen to be explicit in the Helmholtz energy: $A \equiv H - TS$, where $H$ is a Hamiltonian. Similarly, we could utilize $T$ as a main temperature representative with $T/T_\nu$ as a constraint complementary to $NL_\nu$. However, for the gas-liquid systems we could also utilize our approach (28) with $A \equiv H/T - S$. We prefer approach (28) as more symmetric.

The true free energy potential that should be maximized for paths is the probability of these paths $W^{equil}_{[L_0]\to[L_\nu]}$, given by equation (17). However, $T_\nu$, $T$ and $Z^{equil}$ are positive constants and the logarithmic function is the monotonically increasing dependence. Therefore we see that the Helmholtz energy can play the role of the free energy potential that should be minimized.

In Gibbs equilibrium statistical mechanics an equilibrium state is found as a minimum of a free energy potential. For the microcanonical ensemble the free energy potential is the negative entropy $-S \equiv < \ln w_{\{n\}} > \equiv \sum_{\{n\}} w_{\{n\}} \ln w_{\{n\}}$ ; for the canonical ensemble the free energy potential is the Helmholtz energy $A \equiv < H_{\{n\}} > + T < \ln w_{\{n\}} > \equiv \sum_{\{n\}} w_{\{n\}} [H_{\{n\}} + T \ln w_{\{n\}}]$ (for the canonical ensemble the minimization of the Helmholtz free energy is sometimes referred to as a maximization of the entropy with the additional artificial constraint of the equilibrium energy. For the detailed discussion we refer a Reader to section 4 of [10]). For the case of general ensemble in Gibbs equilibrium statistical mechanics the principle of the minimization of the free energy potential always works because this potential is always proportional to the minus logarithm of the probability distribution with the external boundary constraints as constants of proportionality [for detailed discussion see 10]. We see that the same principle is valid and for the case of non-equilibrium mechanics, only now we have to construct the free energy potential not for the states but for the paths. So, for the path microcanonical ensemble the negative dynamical entropy

$$-S \equiv < \ln w_{\{n_0\} \to \{n_v\}} > \equiv \sum_{\{n_0\} \to \{n_v\}} w_{\{n_0\} \to \{n_v\}} \ln w_{\{n_0\} \to \{n_v\}} \tag{29}$$

plays the role of the free energy potential. For the path canonical ensemble the role of the free energy potential is played by the Helmholtz energy

$$A \equiv < H_{\{n_0\} \to \{n_\nu\}} > + T < \ln w_{\{n_0\} \to \{n_\nu\}} > \equiv \sum_{\{n_0\} \to \{n_\nu\}} w_{\{n_0\} \to \{n_\nu\}} [H_{\{n_0\} \to \{n_\nu\}} + T \ln w_{\{n_0\} \to \{n_\nu\}}] \tag{30}$$

where $H_{\{n_0\} \to \{n_\nu\}}$ is the dynamical Hamiltonian which appears in the Boltzmann distribution of probabilities $w_{\{n_0\} \to \{n_\nu\}} = \exp(-H_{\{n_0\} \to \{n_\nu\}} / T) / Z^{equil}$ for the path canonical ensemble. This dynamical Hamiltonian does not correspond to the classical Hamiltonian of states and is determined by the rules of the memory of the process.

For the equilibrium macropath $[L_0^{(0)}] \to [L_\nu^{(0)}]$ the Helmholtz energy is $A_{[L_0^{(0)}] \to [L_0^{(0)}]} \approx -\ln Z^{equil}$ which coincides with the equilibrium Helmholtz energy obtained by the averaging of the ensemble in equilibrium with the SC $P$

$$A^{equil} \equiv \sum_{\{n_0\} \to \{n_\nu\}} \left\{ N L_\nu / T_{L_\nu} + N \left( \sum_{i=1}^{\nu} L_{i-1} \right) / T \right\} w_{\{n_0\} \to \{n_\nu\}}^{equil} - S^{equil} =$$

$$\tag{31}$$

$$= - \sum_{\{n_0\} \to \{n_\nu\}} \left\{ -\ln w_{\{n_0\} \to \{n_\nu\}}^{equil} - N L_\nu / T_{L_\nu} - N \left( \sum_{i=1}^{\nu} L_{i-1} \right) / T \right\} w_{\{n_0\} \to \{n_\nu\}}^{equil} = -\ln Z^{equil} .$$

At the point of the maximum of $W_{[L_0] \to [L_\nu]}^{equil}$ we have

$$\frac{\partial W_{[L_0] \to [L_\nu]}^{equil}}{\partial L_i} [L^{(0)}(t)] = 0 \ \text{ or } \ \frac{\partial \ln W_{[L_0] \to [L_\nu]}^{equil}}{\partial L_i} [L^{(0)}(t)] = 0 \tag{32}$$

(where $i \geq 1$ as we assume $L_0$ not to be variable). For $\ln W_{[L_0] \to [L_\nu]}^{equil} = \ln g_{[L_0] \to [L_\nu]} + A^{equil} - N L_\nu / T_{L_\nu} - N \left( \sum_{i=1}^{\nu} L_{i-1} \right) / T$ we can write that

$$\frac{1}{T} = \frac{\partial \ln g_{[L_0] \to [L_\nu]}}{N \partial L_i} \bigg|_{L^{(0)}(t)} = \frac{\partial \ln g_{[L_0^{(0)}] \to [L_\nu^{(0)}]}}{N \partial L_i^{(0)}}, i = 1, \ldots, \nu - 1$$

and

$$\tag{33}$$

$$\frac{1}{T_\nu} = \frac{\partial \ln g_{[L_0] \to [L_\nu]}}{N \partial L_\nu}\bigg|_{L^{(0)}(t)} = \frac{\partial \ln g_{[L_0^{(0)}] \to [L_\nu^{(0)}]}}{N \partial L_\nu^{(0)}}$$

at the equilibrium macropath. These equations could be used as a definition of the temperature. As both the entropy of a macropath $S_{[L_0] \to [L_\nu]} = \ln g_{[L_0] \to [L_\nu]}$ and the equilibrium entropy $S^{equil} \approx S_{[L_0^{(0)}] \to [L_\nu^{(0)}]}$ have the same functional dependence on $L(t)$ and $L^{(0)}(t)$ respectively, we obtain

$$\frac{1}{T} = \frac{\partial S_{[L_0] \to [L_\nu]}}{N \partial L_i}\bigg|_{L^{(0)}(t)} \approx \frac{\partial S^{equil}}{N \partial L_i^{(0)}}, i = 1, ..., \nu - 1 \text{ and } \frac{1}{T_\nu} = \frac{\partial S_{[L_0] \to [L_\nu]}}{N \partial L_\nu}\bigg|_{L^{(0)}(t)} \approx \frac{\partial S^{equil}}{N \partial L_\nu^{(0)}}. \quad (34)$$

This is an analog of the energy-balance equation - an equation of 'dynamical' balance

$$N\left(\sum_{i=1}^{\nu} dL_{i-1}^{(0)}\right)/T + NdL_\nu^{(0)}/T_\nu = dS^{equil}.$$ This equation could be obtained directly by the

differentiating equation (21) as the logarithm of equation (16).

To find the equilibrium macropath $[L_0^{(0)}] \to [L_\nu^{(0)}]$ we should find when the derivatives of the probability of macrostates (14) (or of the logarithm of this probability) with respect to $L_i$ equals zero (equation (32)). This provides

$$L_i^{(0)} = P^i, \ |\Delta L_i^{(0)}| = (1-P)P^{i-1}, \ i = 1, ..., \nu. \quad (35)$$

as an equation of state.

To investigate the behavior of fluctuations we should find the second derivatives of $\ln W_{[L_0] \to [L_\nu]}^{equil}$:

$$\frac{\partial^2 \ln W_{[L_0] \to [L_\nu]}^{equil}}{\partial L_i \partial L_j}\bigg|_{L^{(0)}(t)} = -\mathrm{K}_{i,j}, \ i, j = 1, ..., \nu, \quad (36)$$

where $K_{i,j}$ is the symmetric matrix of covariance, non-zero elements of which are

$$K_{i,i} = \frac{N}{1-P} \cdot \left\{ \frac{1+P}{P^i} \right\}, \ K_{i,i+1} = K_{i+1,i} = \frac{N}{1-P} \cdot \left\{ -\frac{1}{P^i} \right\}, \ K_{\nu,\nu} = \frac{N}{1-P} \cdot \left\{ \frac{1}{P^\nu} \right\}. \tag{37}$$

For the probability of fluctuations around the equilibrium path we obtain

$$\ln W^{equil}_{[L_0] \to [L_\nu]} = \ln W^{equil}_{[L_0^{(0)}] \to [L_\nu^{(0)}]} - \frac{1}{2} \sum_{i,j=1}^{\nu} (L_i - L_i^{(0)}) K_{i,j} (L_j - L_j^{(0)}) + ... \ \text{or}$$

$$\tag{38}$$

$$W^{equil}_{[L_0] \to [L_\nu]} \propto \exp \left\{ -\frac{1}{2} \sum_{i,j=1}^{\nu} (L_i - L_i^{(0)}) K_{i,j} (L_j - L_j^{(0)}) \right\}.$$

The quadratic form $K_{i,j}$ is positively defined. This can be proved directly. However, we see that it is easy to diagonalize the form $K_{i,j}$. Indeed, transferring to the new coordinates

$$\xi_i = (L_i - L_i^{(0)}) - \frac{1-P^i}{1-P^{i+1}} (L_{i+1} - L_{i+1}^{(0)}), \ i = 1, \dots, \nu - 1; \ \xi_\nu = (L_\nu - L_\nu^{(0)}) \tag{39}$$

we see that the new diagonal quadratic form is

$$\widetilde{K}_{i,i} = \frac{N}{1-P} \cdot \left\{ \frac{1-P^{i+1}}{(1-P^i)P^i} \right\}, \ \widetilde{K}_{\nu,\nu} = \frac{N}{1-P} \cdot \left\{ \frac{1-P}{(1-P^\nu)P^\nu} \right\}, \tag{40}$$

which is positively defined. Also we see that the true order parameters are not $L_i$ but the quantities given by equation (39). For these quantities we have

$$\xi_i = (L_i - L_i^{(0)}) - \frac{1-L_i^{(0)}}{1-L_{i+1}^{(0)}} (L_{i+1} - L_{i+1}^{(0)}), \ i = 1, \dots, \nu - 1; \ \xi_\nu = (L_\nu - L_\nu^{(0)}). \tag{41}$$

In the limit $P \to 1$ (the probability for a fiber to fail at a time-step is much less than unity) we obtain

$$\xi_i = (L_i - L_i^{(0)}) - (L_{i+1} - L_{i+1}^{(0)}) \, , \, i = 1, \ldots, v - 1 ; \; \xi_v = (L_v - L_v^{(0)}) \qquad (42)$$

as expected. Indeed, the process is the first order Markov process and the fluctuations $(L_{i+1} - L_{i+1}^{(0)})$ at the time-step $t_{i+1}$ depend on what the system was at the previous time-step $t_i$. But at the previous time-step the system realized the fluctuations $(L_i - L_i^{(0)})$. Therefore, the new fluctuations depend on the previous fluctuations as on the relative state, from which the new move starts. Similarly to how we moved from the states to the paths, for the order parameters we have to move from the quantities to the changes of the quantities.

So, fluctuations are Gaussian and relative fluctuations are proportional to $1/\sqrt{N}$. Therefore the maximum of the function $W_{[L_0] \to [L_v]}^{equil}$ is indeed very narrow in the thermodynamic limit.

## 6. Conclusions

In this paper we have developed the formalism of non-equilibrium statistical mechanics for non-equilibrium damage phenomena. Far from the state of equilibrium we switched from the states to the paths to base our theory on the most basic quantities which directly determine the probability ensembles. We developed non-equilibrium statistical mechanics for the path ensembles and found the equation of state, the path balance equation, the expression for the dynamical entropy and the free energy potential. Also we showed that the ensemble of systems can be described in terms of the effective temperatures. Although we used the fiber-bundle model with noise and the decay fiber-bundle model to illustrate all concepts developed, we believe that our results have general applicability for other, less simplified damage phenomena.

Another important result of this paper is that we generalized Gibbs principle of the minimization of the free energy potential for path ensembles also, only in this case instead of characteristics of the states we had to move to the dynamical characteristics of the paths.